# Beyond the Gouy-Chapman Model with Heterodyne-Detected Second Harmonic Generation

*Paul E. Ohno,[1] HanByul Chang,[1] Austin P. Spencer,[1] Yangdongling Liu,[1] Mavis D. Boamah,[1] Hong-fei Wang,[2] and Franz M. Geiger*[1]*

[1]Department of Chemistry, Northwestern University, 2145 Sheridan Road, Evanston, IL 60208;

[2]Department of Chemistry and Shanghai Key Laboratory of Molecular Catalysis and Innovative Materials, Fudan University, Shanghai 200433, China

Corresponding Author: geigerf@chem.northwestern.edu

**ABSTRACT.**

We report ionic strength-dependent phase shifts in second harmonic generation (SHG) signals from charged interfaces that verify a recent model in which dispersion between the fundamental and second harmonic beams modulates observed signal intensities. We show how phase information can be used to unambiguously separate the $\chi^{(2)}$ and interfacial potential-dependent $\chi^{(3)}$ terms that contribute to the total signal and provide a path to test primitive ion models and mean field theories for the electrical double layer with experiments to which theory must conform. Finally, we demonstrate the new method on supported lipid bilayers and comment on the ability of our new instrument to identify hyper-Rayleigh scattering contributions to common homodyne SHG measurements in reflection geometries.

**TOC GRAPHIC.**

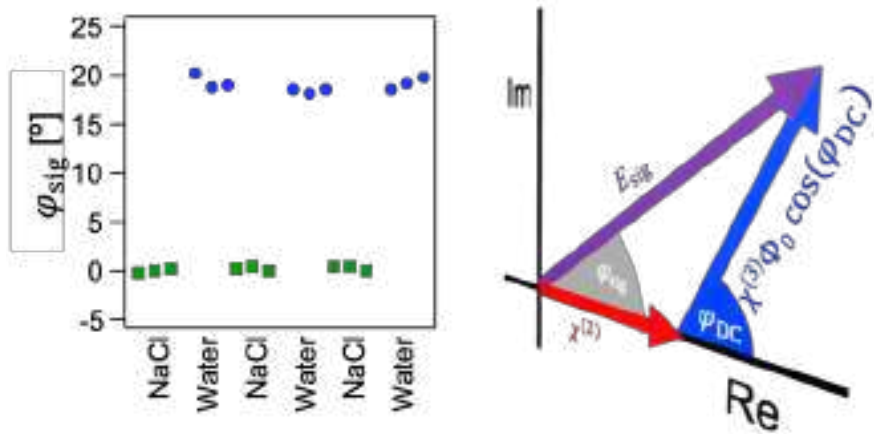

**TEXT.**

The application of second harmonic generation (SHG) and sum frequency generation (SFG) to charged aqueous interfaces has been an area of substantial interest for several decades.[1-7] The field has been greatly influenced by the foundational work of Eisenthal and co-workers,[1] who interpreted the SHG signal generated from the fused silica/water interface, $E_{SHG}$, as consisting of a second-order component, $\chi^{(2)}$, and an interfacial-potential dependent third-order component, $\chi^{(3)}$, using the following model (1):

$$E_{SHG} \propto \chi^{(2)} + \chi^{(3)} \Phi(0) \qquad (1)$$

Here, $\Phi(0)$ is the interfacial potential present at the zero plane of the interface, referenced to zero potential in the bulk solution. The $\chi^{(2)}$ term in eq. 1 originates from molecules that are net oriented at the interface. The interfacial potential-dependent $\chi^{(3)}$ term is present at charged interfaces due to the presence of a static (DC) E-field generated by the surface charge and primarily results from the reorientation and polarization of water molecules in response to the static E-field.[8] Because the penetration of the static E-field from the surface into the aqueous solution depends on the electrostatic screening within the electrical double layer (EDL), the Eisenthal-$\chi^{(3)}$ effect makes SHG and, analogously, SFG, a sensitive probe of interfacial potential and EDL structure.

Many attempts have been made to disentangle the $\chi^{(2)}$ and $\chi^{(3)}$ contributions.[9-16] A recent study of ours identified the $\chi^{(3)}$ contribution to be of bulk origin.[17] Around the same time, Tian and co-workers[6] and Roke and co-workers[7] updated the purely additive model (1)[1,18-21] to account for the optical dispersion between the fundamental and second-harmonic/sum-frequency wavelengths within the interfacial region. For an electrostatic potential exponentially decaying with distance, z, from the interface with a Debye screening length, $\lambda_D$, $\Phi(z) \propto e^{-z\lambda_D^{-1}}$, the interference between signal generated at different depths away from the interface results in the now firmly established model (2):[6-8,17,22-26]

$$E_{SHG} \propto \chi^{(2)} + \chi^{(3)} \Phi(0) \cos(\varphi) e^{i\varphi} \qquad (2)$$

with the phase angle, $\varphi$, of the $\chi^{(3)}$ term taking the exact solution (derivation of this form can be found in the SI, Section 1):[17,22-24,26]

$$\varphi = \arctan(\Delta k_z \, \lambda_D) \qquad (3)$$

where $\Delta k_z$ is the wavevector mismatch of the optical process (calculation of $\Delta k_z$ can be found in the SI, Section 2). We now report phase measurements obtained using a new instrument that we analyze using model (2) so as to unambiguously separate the $\chi^{(2)}$ and interfacial potential-dependent $\chi^{(3)}$ terms. The results provide a path to test primitive ion models and mean field theories for the electrical double layer with experiments to which theory must conform. Moreover, the approach advances the utility of SHG from charged interfaces as an "optical voltmeter".[27] Finally, we demonstrate the new method on supported lipid bilayers and comment on the ability of our new instrument to identify hyper-Rayleigh scattering contributions to common homodyne SHG measurements in reflection geometries.

    The phase angle $\varphi$ in model (2) is not the phase inherent to $\chi^{(3)}$, which varies when $\chi^{(3)}$ is on or near resonance. Rather, $\varphi$ results from the fact that the DC-field induced $\chi^{(3)}$ signal is generated throughout a range of depths away from the interface. For clarity, we will subsequently label this phase angle the DC phase angle, $\varphi_{DC}$, in order to distinguish it from the phase of the overall signal, $\varphi_{\text{sig}}$, measured in the subsequently described experiment using a-quartz against 100 mM NaCl as a reference state. We see from eq. 3 that $\varphi_{DC}$ is a function of the ionic strength of the bulk solution (which determines the Debye screening length) and the wavelengths and angles of the input and output beams. The result of the interference is both a modulation of the

amplitude of the $\chi^{(3)}$ term and a shift in its phase as ionic strength is varied. Simply put, eqs. 2 and 3 show that the "+" in model (1) must be replaced with "+" or "-" or any value in between, depending on the ionic strength of the aqueous solution, as is now done in model (2).

In the case of non-resonant SHG measurements, the inherent phases of $\chi^{(2)}$ and $\chi^{(3)}$ are expected to be purely real, *i.e.* precisely in phase (0°) or out of phase (180°) with respect to the excitation field. However, eq. 2 makes it clear that the overall DC-field induced $\chi^{(3)}$ term can still be phase-shifted relative to the $\chi^{(2)}$ contribution. Our earlier study[17] showed constructive and destructive interference between the surface and bulk terms from the α-quartz/water interface, which, due to the inherent 90° phase shift between surface and bulk terms derived from Maxwell's equations,[28-29] would not be expected according to eq. 1 when $\chi^{(2)}$ and $\chi^{(3)}$ are purely real. This interference was explained by the phase factor included in eq. 2, and the measurements provided experimental evidence for the validity and importance of eqs. 2 and 3. However, at that time, we made no attempt to quantify the phase shift nor deduce what additional information its measurement can provide, which we report here now.

Standard (homodyne) SHG experiments measure only the intensity of the SHG signal, not its phase. Heterodyne-detected SHG (HD-SHG), capable of resolving phase, requires interference between the SHG signal and SHG generated from a reference, called the local oscillator (LO).[30] The phase information encoded is then recovered in the time domain by varying the phase between the signal and LO through the use of a phase shifting unit (PSU).[31] Prior reports of HD-SHG have been largely limited to condensed matter bound by non-dispersive media.[29,32-36] In contrast, determining the phase of the signal generated at a buried interface such as fused silica/water is challenging due to the spatial and temporal dispersion between the fundamental beam and SHG signal as they both propagate away from the interface. This dispersion complicates the generation of a LO that must be collinear and co-temporal with the signal in order for the signal and LO to propagate together and interfere at the detector. The correction of spatial dispersion has been demonstrated through the use of compensating prisms,[37] though this correction does not address temporal dispersion, which represents a significant challenge as even a few mm of glass can be enough to reduce or completely eliminate the temporal overlap.

We earlier demonstrated[38] an SHG phase measurement from the fused silica/water interface by using a hemisphere to avoid refraction of the beams exiting the sample, recollimating the fundamental and SHG beams close to the hemisphere to minimize spatial dispersion, and detecting the signal through a monochromator, which stretches the pulses in time and can make up for some loss of temporal overlap. However, likely because the poor temporal overlap caused by dispersion in the hemisphere was not directly addressed in that study, the efficiency of the interference was low, resulting in a low signal-to-noise ratio and rendering the detection of the small phase shifts expected from changes in $\varphi_{DC}$ difficult or impossible.

In this Letter, we report SHG phase measurements from the fused silica/water interface with a significantly improved signal to noise in order to directly measure phase shifts that occur due to changing EDL thickness. We correct for the temporal dispersion caused by the fused silica hemisphere with a calcite time delay compensator (TDC) and minimize spatial dispersion and chromatic aberration by recollimating with an achromatic off-axis parabolic (OAP) mirror (see SI, Section 3, for calculations of the spatial and temporal dispersion in our setup). We utilize a "sample-first" geometry and generate the LO in a 50 μm z-cut α-quartz plate. With this HD-SHG instrument, we directly measure phase shifts in the SHG signal from the fused silica/water interface at different ionic strengths, as predicted by eq. 2.

The instrument is designed such that homodyne SHG is readily measured in the same setup as HD-SHG through removal of the reference α-quartz crystal (see schematic in Fig. 1). Homodyne SHG control studies measured in this way show the expected quadratic dependence of detected signal intensity on input power as well as the expected narrow bandwidth of the detected signal centered around the second harmonic wavelength of our fundamental beam (SI, Section 4).

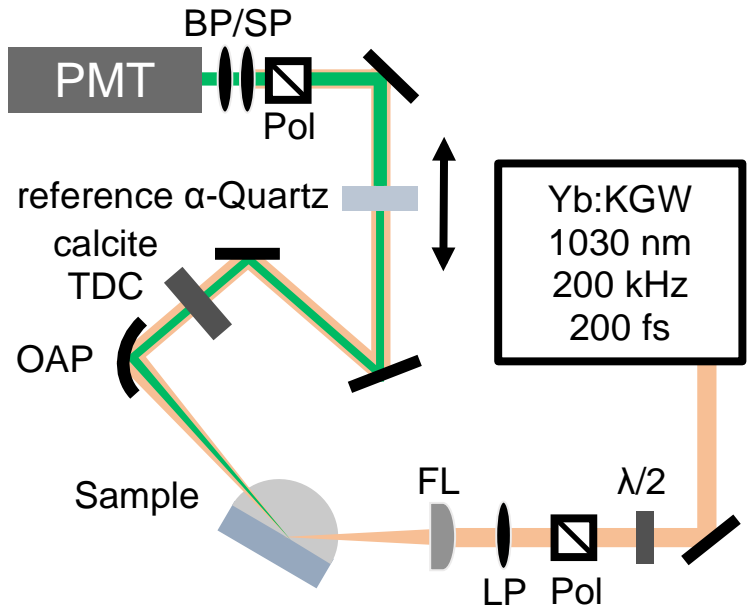

**Figure 1**. **Schematic of PR-SHG instrument**. λ/2 = half-waveplate, Pol = polarizer, LP = long pass filter, FL = focusing lens, OAP = off-axis parabolic mirror, TDC = time delay compensator, SP = short pass filter, BP = bandpass filter, PMT = photomultiplier tube. The reference α-quartz crystal is mounted on a 100 mm translational stage. Part numbers and specifics are provided in SI section 6.

We next show that the HD-SHG instrument yields the expected interference between signal and LO. Fig. 2A shows the homodyne $I_{SHG}$ from the fused silica/water interface is low. Addition of the α-quartz crystal amplifies the $I_{SHG}$ considerably due to the generation of the LO, while translation of the crystal along the beam path generates an interference pattern. Rotating the reference crystal azimuthally by 60°, which changes the phase of the LO by 180°, inverts the interference pattern. Without the calcite TDC plate present and aligned such that it re-overlaps the fundamental and SHG pulses in time prior to their incidence upon the α-quartz crystal, no interference is seen as the crystal is translated, demonstrating the importance of the TDC in the experimental setup.

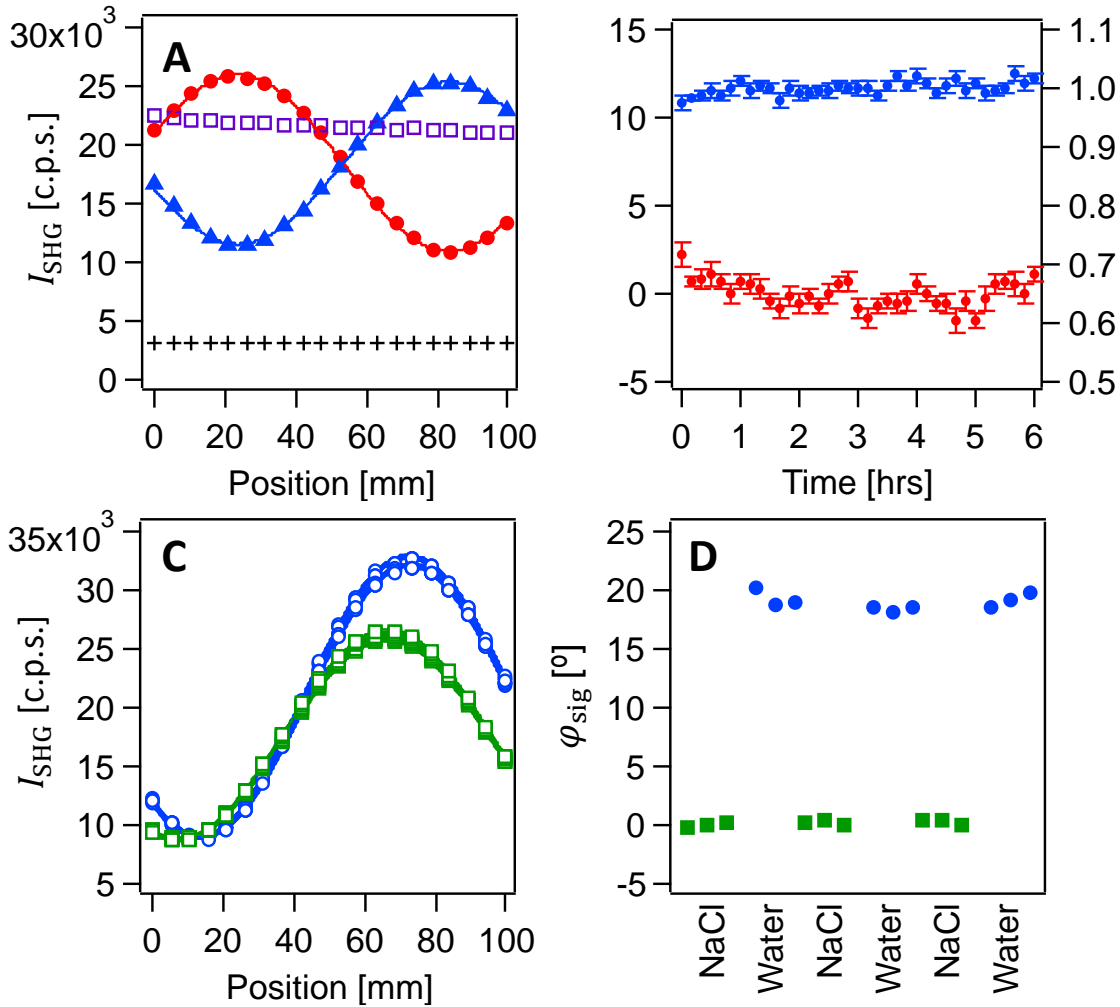

**Figure 2. HD-SHG Measurements. A.** Homodyne SHG measurement (+) and interference patterns (●  and ▲) with fits from the fused silica/100 mM NaCl interface. The phase of the LO for the blue trace has been shifted by 180°. Without the TDC, no interference is seen (□). Fits can be found in Table S2. **B.** Stability of the measurement over time. The phase is shown in red and $A$, proportional to $E_{sig}$, is shown in blue. **C.** Interference patterns and fits from cycling between 100 mM NaCl (□) and 2 μm air-equilibrated water, pH 5.8 (●). **D.** $\varphi_{sig}$ extracted from the fits in (C) shows a reversible phase shift of 19.1 ± 0.4° at pH 5.8.

The phase extracted from the HD-SHG measurements remains stable over the course of hours, as demonstrated in Fig. 2B, most likely due to our chosen collinear geometry as changes in shared optics affect both beam paths equally. However, small shifts in beam pointing or sample position will lead to differing path lengths through the dispersive fused silica hemisphere, causing phase drift. Indeed, we noted a consistent phase drift of ~10° over 2 hours immediately following the hemisphere being clamped to the stage, attributed here to structural relaxation of the flow/optical cell assembly. However, after this initial relaxation period, Fig. 2B shows <±2° phase drifts over 6 hours, with amplitude measurements that fluctuate by ~2%. As individual scans take only 5 minutes, phase measurements taken in succession can be made with a precision of better than 1°.

Equation 2 predicts a difference in the phase of $E_{sig}$ from a fused silica substrate in contact with 2 µM air-equilibrated water vs. 100 mM NaCl, with the exact magnitude of this phase shift determined by the relative amplitudes of the $\chi^{(3)}$ and $\chi^{(2)}$ terms. Though our previous study[17] showed evidence for the existence of this phase shift, we were not able to directly measure its magnitude at that time. With our new HD-SHG setup, we detect a clear phase shift between 2 µM air-equilibrated water relative to 100 mM NaCl, shown in Figs. 2C and 2D (triplicate measurement in three successions). The measurements give $\varphi_{sig,2\mu M} = 19.1 \pm 0.4$ degrees and serve as direct evidence that the phase shift expressed in eq. 2 must be taken into account when SHG and SFG are generated from interfaces at low ionic strengths.

Separation of $\chi^{(2)}$ and $\chi^{(3)}$ contributions from the detected signal intensity in order to use SHG as an optical voltmeter has been a longstanding goal. Previously, without phase information, it was impossible to determine if the observed changes in $E_{sig}$ resulted from changes in the $\chi^{(2)}$ term, the $\chi^{(3)}$ potential-dependent term, or both. We now demonstrate how this goal is experimentally attainable using HD-SHG. Figure 3A shows $\varphi_{sig}$ and $E_{sig}$ extracted from phase measurements like the ones shown in Fig. 2 carried out as a function of ionic strength. $E_{sig}$ initially increases upon addition of salt before decreasing at high ionic strength, consistent with SHG intensity ($I_{SHG}$) measurements reported previously.[26,39] Yet, here we report the amplitude of this response ($E_{sig}$) from the HD-SHG measurement, not merely by square rooting $I_{SHG}$.

The optical process is illustrated in a vector diagram of the signal field in the complex plane, shown in Fig. 3B. The phase, $\varphi_{sig}$, and amplitude of the sample SHG, $E_{sig}$, are the experimental observables extracted from the measured interference patterns and are shown in grey and purple, respectively, corresponding to the symbol colors used in Fig. 3A. Following eq. 2, $E_{sig}$ is modeled to be the sum of a $\chi^{(2)}$ term (shown in red and purely real in our non-resonant experiments), and a $\chi^{(3)}\Phi(0)$ term, shown in blue, whose phase and amplitude are modulated by $\varphi_{DC}$. From trigonometry, we find that

$$\chi^{(3)}\Phi(0) = \frac{\sin(\varphi_{sig})E_{sig}}{\cos(\varphi_{DC})\sin(\varphi_{DC})} \qquad (4A)$$

$$\chi^{(2)} = \cos(\varphi_{sig})E_{sig} - \cos^2(\varphi_{DC})\chi^{(3)}\Phi_0 \qquad (4B)$$

Here, every factor on the right side of eq. 4A is measured ($E_{sig}$, $\varphi_{sig}$) or can be calculated ($\varphi_{DC}$), and, once this is determined, the same holds true for eq. 4B. Thus, with the additional phase information from HD-SHG measurements, we disentangle the $\chi^{(3)}\Phi(0)$ and $\chi^{(2)}$ terms from $E_{sig}$, provided we know the Debye screening length.

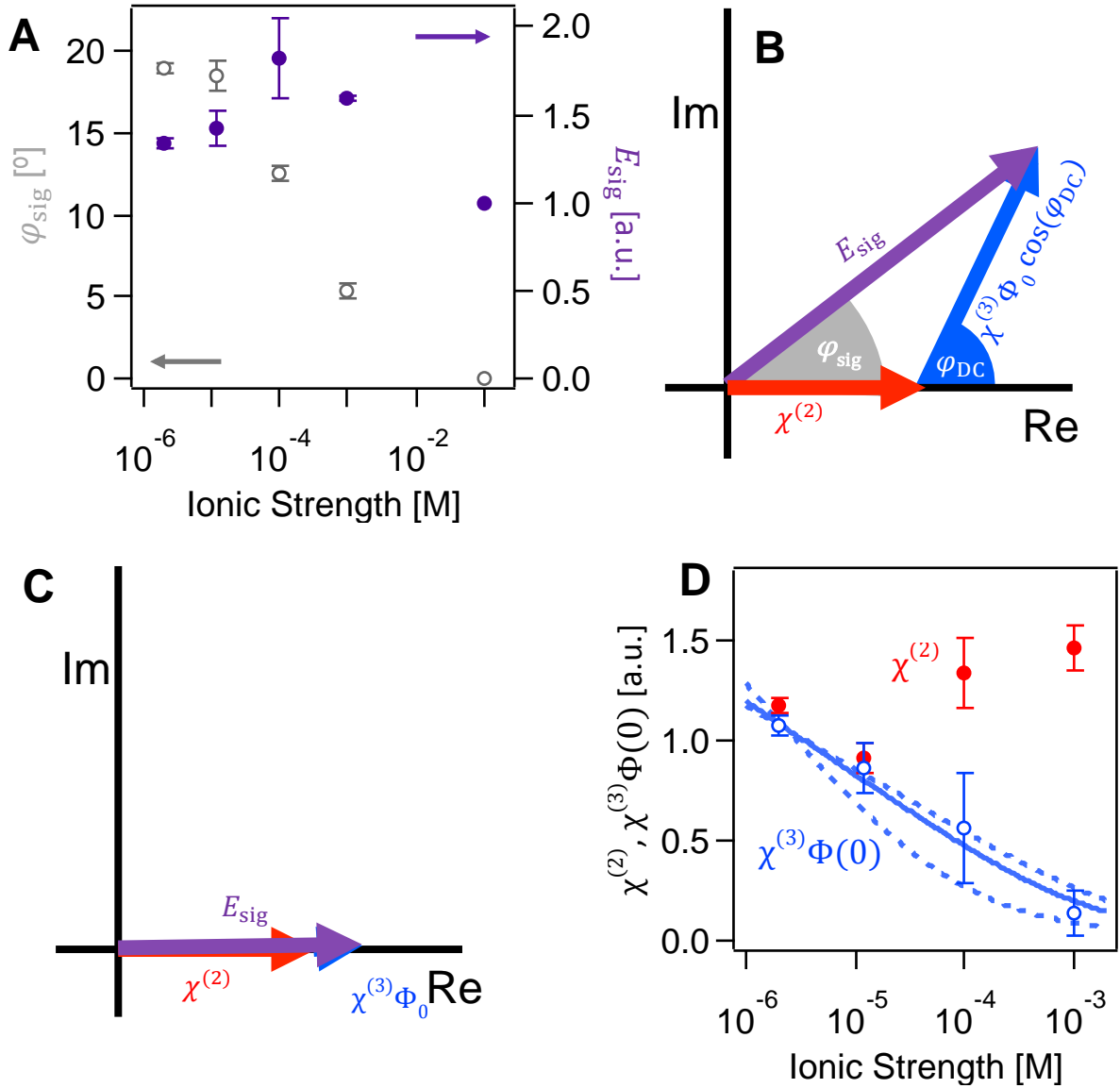

**Figure 3. Separation of $\chi^{(2)}$ and $\chi^{(3)}$ terms. A.** $\varphi_{sig}$ (grey) and $E_{sig}$ (purple) extracted from fits as a function of ionic strength (pH=5.8). Error bars represent the standard deviation from three consecutive measurements. **B.** Graphic representation of real and imaginary components of the signal field at low ionic strength. Because $\chi^{(2)}$ and $\chi^{(3)}$ are themselves purely real, any phase shift can be attributed to $\varphi_{DC}$ according to Eq. 2. **C.** Graphic representation of the signal field at high ionic strength, where the phase shift is minimal and the overall signal remains nearly entirely real. **D.** $\chi^{(2)}$ and $\chi^{(3)}\Phi_0$ calculated from the data in (A) according to Eqs. 4A and 4B, pH=5.8.

Given the experimental[6] and computational[25] evidence that $\chi^{(3)}$ is invariant with the exact nature of the interface and constant across a wide range of aqueous phase conditions up to 100 mM ionic strength, we can interpret the $\chi^{(3)}\Phi(0)$ term as being directly proportional to interfacial potential without having to rely on a model such as Gouy-Chapman or assuming *a priori* that $\chi^{(2)}$ remains constant, as previous studies have posited. Yet, our analysis relies on $\varphi_{DC}$ being large enough such that the overall phase shift, $\varphi_{sig}$, can be reliably detected. Fig. 3C illustrates the case of moderate to high ionic strengths (>~10 mM in our reflection geometry), where the overall signal is nearly entirely real. Considering the noise performance of our instrument illustrated in Fig. 2B, it is not feasible to measure phase shifts of <~1° at this time and to separate the $\chi^{(2)}$ and $\chi^{(3)}$ components at high ionic strength. However, for <~1 mM ionic strength, the phase shift is large enough to measure and separate the terms according to eqs. 4A and 4B. The results of this separation are shown in Fig. 3D, which shows that $\chi^{(3)}\Phi(0)$ decreases significantly with increasing ionic strength, attributable to increased screening within the EDL as the concentration of ions increases. In contrast, $\chi^{(2)}$ remains comparatively constant across 3 orders of magnitude of ionic strength, though changes on the order of ~30% are seen. Note that the magnitude of the uncertainties on the point estimates is largely due to error propagation according to eqs. 4A and 4B. The best fit of $\chi^{(3)}\Phi(0)$ with the Gouy-Chapman model yields a surface charge density of -0.0024(18) C/m$^2$ and is represented by the blue line in Fig. 3D, which agrees well with the measured data. As expected, the charge density, for pH 5.8, is smaller than what has been published for pH 7,[12,20] given the point of zero charge for fused silica is ~2.5.[40]

Our findings support the conclusion that we have successfully separated out the $\chi^{(3)}\Phi(0)$ term without relying on the Gouy-Chapman model. Measurements of $E_{sig}$ referenced to the $E_{sig}$ generated from an interface with known $\chi^{(2)}$ value, such as α-quartz, are now needed to obtain $\Phi(0)$ absolutely, without the use of Gouy-Chapman theory or any other model for the interfacial potential. The method therefore opens the possibility to test primitive ion models or mean field theories for aqueous interfaces directly, and without externally applied labels.

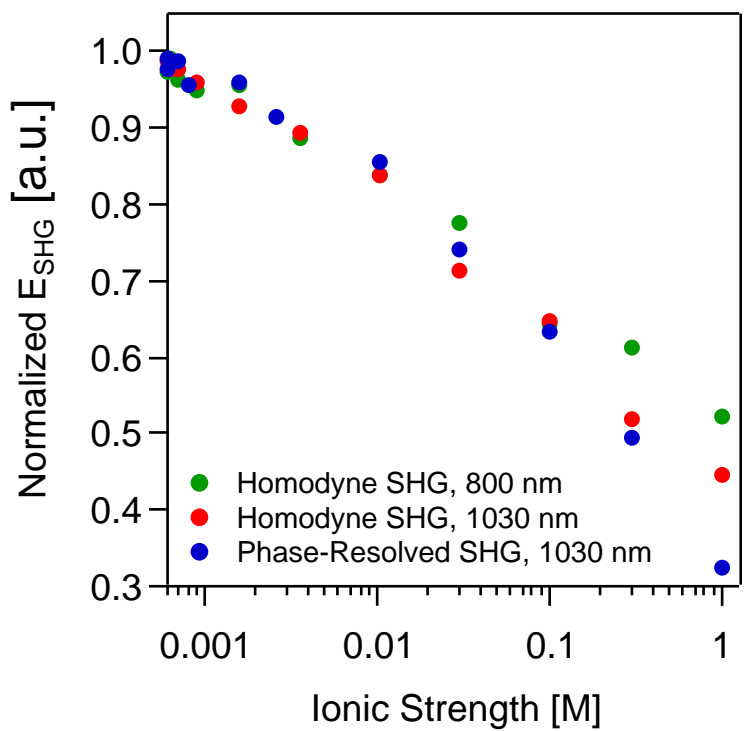

**Figure 4. Homodyne- and HD- SHG Comparison.** Comparison of $E_{sig}$ derived from homodyne and HD-SHG measurements from a supported lipid bilayer/water interface as a function of ionic strength. The greater amplitude detected by the homodyne measurements at high ionic strength is consistent with the presence of HRS.

On a slightly different albeit highly relevant note, we mention at this point a recent investigation by Dreier *et al.*[41] into the surface potential of charged lipid monolayer-water interfaces in which seemingly different results were obtained between homodyne non-resonant SHG measurements and chemically-specific SFG measurements of the OH stretching region. Part of this difference was attributed to hyper-Raleigh scattering (HRS) contributions to the detected SHG intensity. As HRS is produced incoherently, any HRS emitted from the sample – while it may be present in the homodyne measurements – would not contribute to the interference from which the amplitude and phase of the signal SHG are extracted in HD-SHG. Thus, our HD-SHG measurements are free from any potential convolution with HRS. Indeed, our HD-SHG apparatus should be ideal for separating out any possible contributions of HRS to the SHG signal generated from any interface. To test this idea, we obtained $E_{sig}$ from supported lipid bilayer/water interfaces as a function of ionic strength measured with different instruments (Fig. 4). In addition to the HD-SHG measurements that yield $E_{sig}$ by phase-referencing, we obtained $E_{sig}$ by square rooting homodyne-detected $I_{SHG}$ signals with the same laser system as well as with an 800 nm Ti:Sapphire oscillator system described previously.[17] The three measurements track each other closely at low ionic strength (<100 mM), while at high ionic strength (approaching 1 M), both homodyne measurements indicate a higher $E_{sig} = \sqrt{I_{SHG}}$ than the $E_{sig}$ obtained from the HD-SHG measurement through model 2 ($E_{sig} \propto A$ from eqs. 5 and 6, *vide infra*). Additionally, the 400 nm homodyne measurement indicates more signal than the 515 nm measurement, which would be expected as shorter wavelengths produce a greater HRS intensity.[42] We caution that our observations are not conclusive evidence for the presence of HRS in homodyne SHG in reflection geometries such as the ones employed here and elsewhere, yet this result is consistent with its presence and will be the subject of future studies.

In conclusion, we have demonstrated an experimental apparatus capable of measuring the phase of SHG signals from buried interfaces. We used this apparatus to measure the phase shift in SHG generated at the fused silica/water interface as a function of ionic strength, a direct result of charge screening compressing the width of the EDL as ionic strength increases. Furthermore, we showed how this additional phase information can be used to unambiguously separate the $\chi^{(2)}$ and $\chi^{(3)}$ contributions to detected SHG from charged interfaces, a longstanding goal in the field. While model (2) was successfully applied to the experimental data, we caution that the analysis relies on three assumptions: a) the nonlinear optical signal recorded at the detector consists of only second and third order terms, b) the surface potential decays exponentially with distance, and c) Debye-Hückel theory is applicable for the Debye length.

We envision several avenues of future study in which the instrument described here can be impactful. In one avenue, determination of the absolute magnitude of the $\chi^{(2)}$ and $\chi^{(3)}$ components could be made through comparison with a known absolute reference, similar to the approach used in heterodyne SFG spectroscopy. This comparison, in combination with the isolation of the $\chi^{(3)}$ contribution described here, would allow for a direct, optical quantification of the surface potential at oxide/water interfaces, without relying on the Gouy-Chapman model or the assumption that $\chi^{(2)}$ remains constant. In another avenue, we foresee coupling the phase measurements described in this study with a system containing an electrode under potential control. By controlling the potential and thus the magnitude of the $\chi^{(3)}$ contribution, the comparison of observed phase shifts with the phase shifts predicted by model (2) would open an experimental window into how accurate theoretical (atomistic or coarse grain simulations) and model (primitive ions, continuum models or mean field theory) predictions of the EDL are. Despite our advance, we caution that physics and chemistry not described in model (2) may contribute to the signal generation process in ways that remain to be uncovered.

**Experimental.**

A detailed description of the optical setup, as well as sample and solution preparation can be found in the SI, Section 6. HD-SHG requires interference between the sample SHG signal and the LO generated in an α-quartz crystal. The detected total signal intensity, $I_{SHG}$, produced by the coherent interference between the signal and LO is governed by the following equation:[31]

$$I_{SHG} \propto \left|E_{sig} + E_{LO}\right|^2 = \left|E_{sig}\right|^2 + |E_{LO}|^2 + 2E_{sig}E_{LO}\alpha\cos(\varphi_{sig} - \varphi_{LO} + \varphi_{PSU}) \qquad (5)$$

Where α is the overlap parameter that represents the degree of spatial and temporal overlap between the two beams, $\varphi_{sig}$ and $\varphi_{LO}$ represent the phases of the signal and LO, respectively, and $\varphi_{PSU}$ represents the additional phase shift introduced by the PSU. We vary $\varphi_{PSU}$ by translating the reference α-quartz crystal along the beam path, taking advantage of the slight optical dispersion in air, according to the following equation:[38] $l_0 = \lambda/2\Delta n$, where $l_0$ is the translation distance required for one period of oscillation, $\lambda$ is the fundamental wavelength, and $\Delta n$ is the difference in refractive index between the fundamental and SHG wavelengths. Using data from the literature[43] for the refractive index of our 1030 and 515 nm beams in air, we calculate $l_0$ to be ~114 mm. Thus, with a 100 mm translational stage, we are able obtain an interference pattern of just less than one full period of oscillation with our PSU. The time required for each scan depends on the number of points the 100 mm range of the stage is divided into as well as the length of acquisition at each point. A typical scan of 20 points at 10 seconds per point took ~5 minutes.

Scanning the position of the α-quartz crystal shifts the phase between the signal and LO and results in an interference pattern in $I_{SHG}$ as a function of stage position. We fit the interference pattern to the following equation:

$$I_{SHG} = I_0 + A\cos(fx + \varphi_{\text{fit}}) \qquad \#(6)$$

where $x$ is the stage position and $I_0, A, f,$ and $\varphi_{\text{fit}}$ are parameters free to be optimized. The fitting is carried out using SciPy in JupyterLab in two sequential steps. First, each individual scan from a dataset is fit to eq. 6 with every parameter free to be optimized. As the phase of cosine functions can only be rigorously compared within a set of cosines with precisely the same frequency, the patterns are then fit a second time with $f$ held at the average of all the $f$ values from the data set, and these final parameters are subsequently analyzed. As the signal to noise ratio of the interference patterns is high, the difference in the relevant parameters between these two steps is generally < ~0.5%.

By comparing eqs. 5 and 6, it can be seen that $E_{sig} \propto A$. We do not attempt to deduce the absolute phase of the SHG signal and only interpret its changes in phase. Because $\varphi_{LO}$ is constant and $\varphi_{PSU}$ is varied in the same manner in each scan, changes in $\varphi_{\text{fit}}$ must originate from changes in $\varphi_{sig}$, i.e. $\Delta\varphi_{fit} = \Delta\varphi_{sig}$. To move from $\Delta\varphi_{sig}$ to $\varphi_{sig}$, the phase of the sample under a specific condition must be known (or assumed). We assume that at 100 mM NaCl $\varphi_{sig} = 0°$ as $\chi^{(2)}$ and $\chi^{(3)}$ are purely real and at 100 mM NaCl the Debye length is sufficiently short such that $\varphi_{DC}$ is near 0°. The sign of $\varphi$ depends on the sign convention of the z-axis, a point of disagreement in the literature.[8] We assume $\varphi_{DC}$ is positive, so we use the absolute value of $\varphi$ extracted from the fits. We additionally place $\chi^{(2)}$ and $\chi^{(3)}\Phi(0)$ on the positive real axis of the complex plane, which places $E_{sig}$ in the upper right quadrant in an Argand diagram.

## ASSOCIATED CONTENT

**Supporting Information:** (1) Derivation of eqs. 2 and 3 (2) Calculation of wavevector mismatch (3) Calculation of spatial and temporal overlap (4) SHG control studies (5) Fit parameter from Fig. 2A (6) Experimental Setup

## AUTHOR INFORMATION

**Corresponding author:**

Email: geigerf@chem.northwestern.edu

**ORCID:**

Paul E. Ohno: 0000-0003-4192-1888

HanByul Chang: 0000-0003-0270-7336

Austin P. Spencer: 0000-0003-4043-2062

Yangdongling Liu: 0000-0003-2750-9360

Hong-fei Wang: 0000-0001-8238-1641

Franz M. Geiger: 0000-0001-8569-4045

**Author Contributions.** PEO and FMG conceived of the idea. PEO, APS, and FMG designed the instrument. PEO and HBC collected the data. PEO, HBC, APS, YL, MDB, HFW, and FMG analyzed the data. The manuscript was written with substantial input from all authors.

**Notes.** The authors declare no competing financial interests.

**ACKNOWLEDGMENTS**

This work was supported by the US National Science Foundation (NSF) under its graduate fellowship research program (GRFP) award to PEO. PEO also acknowledges support from the Northwestern University Presidential Fellowship. HFW gratefully acknowledges support from the Shanghai Municipal Science and Technology Commission (Project No. 16DZ2270100) and Fudan University. FMG gratefully acknowledges support from the NSF through award number CHE-1464916. This work was also supported by the NSF under the Center for Sustainable Nanotechnology, Grant No. CHE-1503408. FMG acknowledges a Friedrich Wilhelm Bessel Award from the Alexander von Humboldt Foundation.

*Supporting Information for*

# Beyond the Gouy-Chapman Model with Heterodyne-Detected Second Harmonic Generation

*Paul E. Ohno,[1] HanByul Chang,[1] Austin P. Spencer,[1] Yangdongling Liu,[1] Mavis D. Boamah,[1] Hong-fei Wang,[2] and Franz M. Geiger\*[1]*

[1]Department of Chemistry, Northwestern University, 2145 Sheridan Road, Evanston, IL 60208;

[2]Department of Chemistry and Shanghai Key Laboratory of Molecular Catalysis and Innovative Materials, Fudan University, Shanghai 200433, China

# 1. Derivation of Eq. 2 and 3.

Eq. 2 is often written:

$$E_{SHG} \propto \chi^{(2)} + \frac{\kappa}{\kappa - i\Delta k_z}\chi^{(3)}\Phi_0 \quad \text{(S3)}$$

where $\kappa = \lambda_D^{-1}$. The derivation can be found for instance in Roke et al.[1] Explicitly separating the real and imaginary parts of the correction factor:

$$\frac{\kappa}{\kappa - i\Delta k_z} = \frac{\kappa^2}{\kappa^2 + \Delta k_z^2} + i\frac{\kappa\Delta k_z}{\kappa^2 + \Delta k_z^2} \quad \text{(S4)}$$

Writing it in polar form gives:

$$\frac{\kappa^2}{\kappa^2 + \Delta k_z^2} + i\frac{\kappa\Delta k_z}{\kappa^2 + \Delta k_z^2} = \sqrt{\frac{\kappa^4 + \kappa^2\Delta k_z^2}{(\kappa^2 + \Delta k_z^2)^2}}e^{i\varphi} = \sqrt{\frac{\kappa^2(\kappa^2 + \Delta k_z^2)}{(\kappa^2 + \Delta k_z^2)^2}}e^{i\varphi} = \frac{\kappa}{\sqrt{\kappa^2 + \Delta k_z^2}}e^{i\varphi} \quad \text{(S5)}$$

where:

$$\varphi = \arctan\left(\frac{\frac{\kappa\Delta k_z}{\kappa^2 + \Delta k_z^2}}{\frac{\kappa^2}{\kappa^2 + \Delta k_z^2}}\right) = \arctan\left(\frac{\Delta k_z}{\kappa}\right) = \arctan(\Delta k_z \lambda_D) \quad \text{(S6)}$$

Finally, from the following triangle:

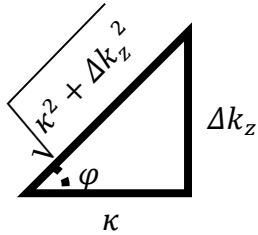

we find that:

$$\frac{\kappa}{\sqrt{\kappa^2 + \Delta k_z^2}} = \cos\varphi \quad \text{(S7)}$$

## 2. Wavevector Mismatch

For SFG/SHG in the reflection geometry as defined above where the $\chi^{(3)}$ process is occurring in Medium II, $\Delta k_z$ is given by the following equations:[1-3]

$$\Delta k_z = |k_{1z,II} + k_{2z,II} - k_{0z,II}| \tag{S1}$$

and $\theta_{i,II}$ can be determined by incident angles and the refractive indices of Media I and II. With Medium I = fused silica, Medium II = water, $\theta_{1,I} = \theta_{2,I} = 60°$, $n_{water}$ @ 1030 nm = 1.3247; $n_{water}$ @ 515 nm = 1.3357; $n_{fused\ silica}$ @ 1030 nm = 1.4500; $n_{fused\ silica}$ @ 515 nm = 1.4615; angle = 60 degrees in fused silica, we find $\Delta k_z = 1.1 \times 10^7$ m$^{-1}$.

**Figure S1. Diagram of Wavevector Mismatch.**

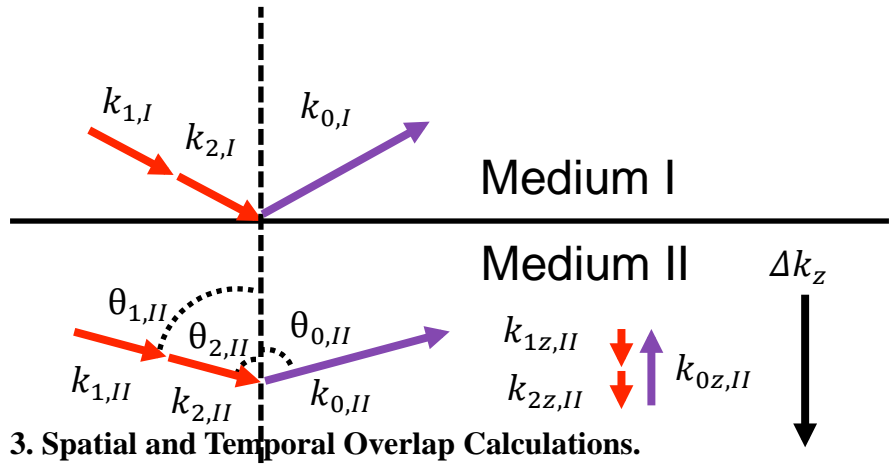

## 3. Spatial and Temporal Overlap Calculations.

### A. Temporal Overlap

The refractive ($n$) and group ($n_g$) indices of the fundamental and SHG wavelengths are shown in **Table S1**.[4] From the formula of the group velocity of a pulse, $v_g = c/n_g$, the time difference per unit length between the two pulses: $\Delta\left(\frac{1}{v_g}\right) = \frac{1}{c}(n_{g,2\omega} - n_{g,\omega}) = 834$ fs/cm. With the common path length in the fused silica hemisphere of 1.27 cm, we predict upon exiting the hemisphere the fundamental pulse will lead the SHG pulse by **1060 fs**. As our pulse length is ~200 fs, this necessitates the use of the calcite time delay compensator.

| Wavelength (nm) | Refractive Index ($n$) | Group Index ($n_g$) |
|---|---|---|
| 515 | 1.4615 | 1.4877 |
| 1030 | 1.4500 | 1.4627 |

**Table S1**. Group and refractive indices of fused silica.

## B. Spatial Overlap

The approximately Gaussian fundamental beam has a $1/e^2$ diameter of 4.5 mm. We focus with an $f = 7.5$ cm lens, and the fundamental and SHG are recollimated using a $f = 10.16$ cm silver OAP, at which distance the fundamental beam diameter = 6 mm. The diameter of the sample SHG beam is theoretically $\frac{1}{\sqrt{2}} \times 6 = 4.2$ mm.

The SHG output angle, $\theta_{2\omega}$, for co-propagating, reflected beams is:[5]

$$n_{2\omega} \sin \theta_{2\omega} = n_\omega \sin \theta_\omega \tag{S1}$$

where $\theta_\omega$ is the input angle. Using the values from **Table S1** and an input angle of 60°, the calculated SHG output angle is 59.2°, for a separation of 0.8° between the fundamental and SHG beams. Assuming the incoming and outgoing beams hit the hemisphere normal, the center of the two beams will be offset by 1.4 mm at the recollimating mirror, and will propagate collinearly to the α-quartz crystal, where the fundamental will produce the LO beam with a diameter of 4.2 mm.

The 2D intensity profile of one beam with a diameter of 4.2 mm in diameter can be represented as:

$$I \propto e^{\frac{-2(x^2+y^2)}{width^2}} \tag{S2}$$

where $width$ = the $1/e^2$ radius, or 2.1 mm in our case. We then calculate the overlap efficiency between the two beams as the integral of the product of the two offset Gaussians normalized by the integral of the product of the two Gaussians with no offset:

$$\alpha = \frac{\iint e^{\left(\frac{-2(x^2+y^2)}{width^2}\right)} e^{\left(\frac{-2((x-x_0)^2+y^2)}{width^2}\right)} dxdy}{\iint e^{\left(\frac{-2(x^2+y^2)}{width^2}\right)} e^{\left(\frac{-2(x^2+y^2)}{width^2}\right)} dxdy} \tag{S3}$$

where $x_0$ = the offset. We numerically solve the integral above and find the overlap efficiency with an offset of 1.4 mm = **0.64**. Thus, despite the separation of the two beams, we still expect appreciable interference, even without the use of spatial compensating optics.

## 4. SHG Control Studies

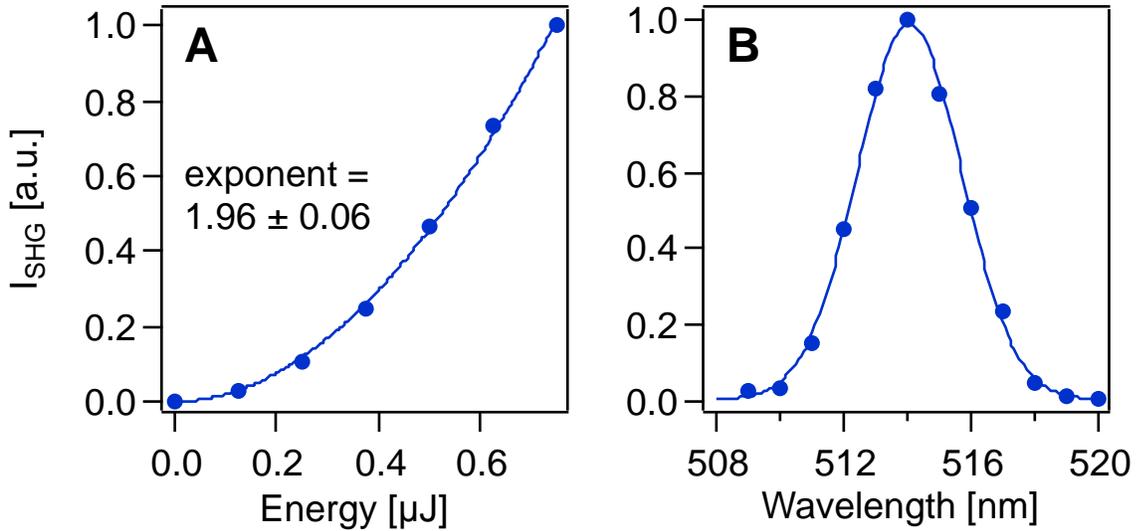

**Figure S2. A.** $I_{SHG}$ as a function of pulse input energy from the fused silica/pure water interface with the $S_{in}$,$P_{out}$ polarization combination and power function fit, which shows the expected quadratic dependence up to 0.75 µJ = 150 mW. In experiments the power was kept below 0.4 µJ, well within the quadratic regime. **B.** Bandwidth study shows a Gaussian centered at 514 nm.

## 5. Fit Parameters from Fig 2A.

| Scan | $y_0$ | $A$ | $f$ | $\varphi(°)$ |
|---|---|---|---|---|
| α-quartz (red) | 18460±30 | 6900±40 | 0.0525 | -70.7±0.4 |
| α-quartz-60° (blue) | 18460±30 | 7500±50 | 0.0525 | 108.7± 0.4 |

**Table S2.** To calculate the overlap parameter α from equation 3, we assume $E_{sig}$ is constant with and without the reference. $I_{sig}$ without the reference (Fig 2A, black) was 3100 c.p.s., so $E_{sig} = \sqrt{3100} = 56$. From Eqs. 3 and 4, $E_{LO} = \sqrt{y_0 - I_{sig}} = 124$. $\alpha = \frac{A}{2E_{sig}E_{LO}} \approx 0.5$, slightly lower than the estimate calculated by the spatial separation of the two beams, 0.64 (Section 3). This lower value is likely because the time mismatch is not perfectly compensated by the TDC.

## 6. Experimental Setup

**A. Optical Setup.** PR-SHG measurements were performed using 200 fs pulses centered at 1030 nm at a repetition rate of 200 kHz (Light Conversion, Pharos). The incident pulse energy during experiments was kept below 0.4 µJ (80 mW), well within the regime where the quadratic dependence of $I_{SHG}$ on input power was confirmed (SI, Fig. S2A). Prior to the sample stage, the input power and polarization were controlled with a half-waveplate (Thorlabs, AHWP05M-980) and Glan-Taylor polarizer (Thorlabs, GT-10), filtered with a long pass filter to remove any residual SHG, and focused into the sample with an f = 7.5 cm lens (Thorlabs, LA1608) with an angle of 60°±1°. The reflected fundamental and generated sample second harmonic were recollimated using an off-axis parabolic mirror (Thorlabs, MPD149-P01) to limit chromatic aberration and avoid further temporal dispersion. The two orthogonally-polarized pulses were then re-overlapped in time with a calcite TDC (Newlight Photonics, CAL12200-A). See SI, Note S2 for a calculation of the amount of spatial and temporal overlap expected of the SHG and fundamental pulses after they exit the hemisphere (64% efficiency). The newly-overlapping pulses were then passed through a 50 μm thick z-cut α-quartz window (Precision Micro-Optics, PWQB-368252) to generate the LO. Azimuthal rotation of the α-quartz was used to tune the intensity of the LO. The reference α-quartz window position along the beam path was controlled by a motorized translational stage (Standa, 8MT193-100). Following LO generation, the pulses were then passed through a second Glan-Taylor polarizer to select the output polarization, a short-pass filter to remove the fundamental beam (Edmund Optics, 45-646), and a bandpass filter (Thorlabs, FBH520-40) to isolate the SHG light prior to incidence upon the detector. For the bandwidth measurement, a monochromator (Dynasil, MC1-02) was inserted into the beam path prior to the detector. Detection was achieved with a photomultiplier tube (Hamamatsu, R585 and H8259-01) in conjunction with a gated photon counter and preamplifier (Stanford Research Systems, SR400 and SR445A) with a gate width of 50 ns. Spurious counts were subtracted from the detected signal intensity using the second channel of the SR400 with the gate offset from the laser pulses. Though we limit our measurements to the $S_{in}$, $P_{out}$ polarization combination, we note that this experimental geometry could be extended to any polarization combination, including $P_{in}$, $P_{out}$, using selective half-wave plates, as demonstrated recently in an SHG microscope.[6]

**B. Sample and Solution Preparation.** One inch diameter fused silica hemispheres (ISP Optics) were clamped to the aluminum exterior of a home-built flow cell. The wetted interior of the flow cell is PTFE and is sealed to the hemisphere with a fluoroelastomer o-ring. Prior to measurements, the hemispheres were treated with NoChromix for 1 hour, sonicated in methanol for 5 minutes, sonicated in water for 5 minutes, rinsed copiously with water, dried, plasma cleaned (Harrick Plasma) for 1 minute, and immediately stored in pure water prior to the start of measurements. The solutions were prepared with Millipore water (18.2 MΩ*cm) and NaCl was procured from Sigma Aldrich (#746398, >99%). All solutions were equilibrated overnight with atmospheric $CO_2$ and were thus at pH 5.7. This $CO_2$-equilibrated water is assumed to have a total ionic strength of 2 µM, consistent with our prior measurements.[7-8] To minimize previously reported hysteresis effects,[9-11] we cycled the solution in the sample cell between pure water and 100 mM NaCl prior to the measurements reported here. Flow conditions in the cell are characterized by low (<1) sheer rates.

**C. Lipid Bilayer Preparation.** For the HD-SHG/homodyne SHG comparison of the lipid membrane/water interfaces, POPC in chloroform was purchased from Avanti Polar Lipids. Rough LPS (rLPS) from *Salmonella enterica* serotype minnesota Re 595 (Re mutant) was purchased from Sigma Aldrich as lyophilized powder and suspended in neat chloroform (B&J Brand, HPLC grade) to 2 mg mL$^{-1}$ by sonicating for at least 20 minutes. Following our previously published work,[12] we prepared lipid bilayers from unilamellar vesicles containing 80% POPC and 20% rLPS (mass ratio). POPC and rLPS were mixed to desired ratios in a glass vial, dried under $N_2$ gas, and then dried in vacuum for at least 6 hours to remove excess chloroform. The dried vesicles were then rehydrated in 1 mL 1 mM NaCl:2 mM HEPES for at least an hour, vortexing occasionally to aid in hydration. Unilamellar lipid vesicles (~100 nm) were prepared by extrusion though a 0.05 µm membrane filter (Avanti, 610000) in 1 mM NaCl:2 mM HEPES at pH 7.4 after 4 cycles of freeze-thaw, 5 minutes each. The vesicles were diluted to 0.5 mg mL$^{-1}$ with appropriate salt concentration for a final buffer concentration of 150 mM NaCl:2 mM HEPES at pH 7.4 immediately before forming supported lipid bilayers using the vesicle fusion method.[13] Following our previously reported procedures,[12] 150 mM NaCl:2 mM HEPES at pH 7.4 (without vesicles) was introduced into the flow cell and the SHG response was recorded until a steady signal was attained for at least 10 minutes. Next, extruded vesicles at a concentration of 0.5 mg mL$^{-1}$ in 150 mM NaCl:2 mM HEPES at pH 7.4 were introduced into the cell at a flow rate of 2 mL min$^{-1}$ for two minutes for a total volume of 4 mL. The vesicles were then allowed to self-assemble into a lipid bilayer on the fused silica substrate for at least 30 minutes. Following bilayer formation, a solution of 150 mM NaCl:2 mM HEPES at pH 7.4 buffer was flushed through the cell at a rate of 2 mL min$^{-1}$ for 10 minutes to remove excess vesicles before starting any experiments. To prepare the NaCl solutions, 2 mM HEPES (Fisher Scientific, powder, purity ≥99%) was dissolved into ultrapure water along with the varying NaCl concentrations used in the experiments. The buffer solutions were then pH-adjusted to pH 7.4 using small aliquots of dilute HCl and NaOH.